\documentclass{amsart}
\usepackage{amsfonts}
\usepackage{graphicx}
\usepackage{hyperref}
\usepackage{multirow}
\usepackage[table]{xcolor}
\newcolumntype{R}[1]{>{\raggedleft\arraybackslash }b{#1}}
\newcolumntype{L}[1]{>{\raggedright\arraybackslash }b{#1}}
\newcolumntype{C}[1]{>{\centering\arraybackslash }b{#1}}

\newtheorem{theorem}{Theorem}[section]
\theoremstyle{plain}

\newtheorem{remark}{Remark}

\numberwithin{equation}{section}

\begin{document}

\begin{center}
\huge Asymptotic confidence bands for copulas based on the transformation kernel estimator
\end{center}
\bigskip
\begin{center}
\large  Diam Ba$^{\dag}$, Cheikh Tidiane Seck$^{\dag\dag}$ and Gane Samb LO$^{\dag\dag\dag}$
\end{center}
\bigskip
\begin{center}
\Large Presented by Mik\`{o}s Cs\"{o}rg\H{o}, FRSC.
\end{center}

\bigskip

\noindent \textbf{Abstract}. In this paper we establish asymptotic simultaneous confidence bands for the transformation kernel estimator of copulas introduced in Omelka et al.(2009). To this aim, we prove a uniform in bandwidth law of the iterated logarithm for the maximal deviation of this estimator from its expectation, under smoothness conditions on the copula function. We also study the bias, which tends asymptotically and uniformly to zero with the same precise rate. Some simulation experiments are finally provided to support our results. \\

\noindent \textit{AMS 2010 Classifications Subjects} : Primary 62G05, 62G07; Secondary 60F12, 62G20.\\

\section{Introduction}
Let $(X_1,Y_1),...,(X_n,Y_n)$ be an independent and identically distributed sample of the bivariate random vector $(X,Y)$ with joint cumulative distribution function $H$ and marginal distribution functions $F$ and $G$. Let $F_n$ and $G_n$ be the empirical marginal cumulative distribution functions and let $K(\cdot,\cdot)$ represent a multiplicative kernel integral $K(x,y)=K(x)K(y)$. Then, the transformation kernel estimator of copulas suggested in \cite{r6} is defined as follows :
\begin{equation}\label{ee1}
\hat{C}_{n,h}^{(T)}(u,v)=\frac{1}{n}\sum_{i=1}^{n}K\left(\frac{\phi^{-1}(u)-\phi^{-1}(\hat{U}_i)}{h} \right)K\left(\frac{\phi^{-1}(v)-\phi^{-1}(\hat{V}_i)}{h} \right),
\end{equation}
where $\phi$ is a given distribution function and $\hat{U}_i=\frac{n}{n+1}F_n(X_i)$, $\hat{V}_i=\frac{n}{n+1}G_n(Y_i)$. Taking $\phi$ equal to the standard Gaussian distribution leads to the Probit transformation proposed by Marron and Ruppert \cite{r5}  (1994). Furthermore, this estimator presents a great advantage as it does not depend on the marginal distributions.
We shall consider in this paper a general bandwidth $h$ that may depend either on the sample data or/and the location $(u,v)$. It is found in the literature that, for practical use, the most interesting choice of the bandwidth for kernel distribution function estimation is the data-driven method (see, e.g., Altman and L\'eger \cite{r0} (1995)).
  
\section{Main results and Simulation}\label{ssec2}

\subsection{Results}

Here, we state our theoretical results in three theorems. The first theorem gives the uniform in bandwidth rate of convergence for the maximal deviation of the estimator \eqref{ee1} from its expectation. The second theorem handles the  bias, while the third  theorem provides optimal asymptotic simultaneous  confidence bands for the copula function $C(u,v)$ defined as,
$$ C(u,v)=\mathbb{P}(U\leq u,V\leq v)\qquad\text{for all}\; 0\leq u,v\leq 1,$$
where $U$ and $V$ are $(0,1)-$uniform random variables.

\begin{theorem}\label{tt1}
Suppose that the copula function $C(u,v)$ has bounded first-order partial derivatives on $(0, 1)^2$ and the function $\phi$ admits a bounded derivative $\phi'$. Then, for any sequence of positive constants $(b_n)_{n\geq 1}$ satisfying $0<b_n<1, b_n\rightarrow 0$, $b_n\geq (\log n)^{-1}$, and for some $c>0$, we have almost surely 
\begin{equation}\label{ee2}
\limsup_{n\rightarrow\infty}\left\{R_n\sup_{\frac{c\log n}{n}\le h\le b_n}\sup_{(u,v)\in(0,1)^2}\left|\hat{C}_{n,h}^{(T)}(u,v)-\mathbb{E}\hat{C}_{n,h}^{(T)}(u,v)\right|\right\}=A(c),
\end{equation}
where $A(c)$ is a positive constant less than or equal to 3, and $R_n =\left(\frac{n}{2\log\log n}\right)^{1/2}$.
\end{theorem}


\begin{theorem} \label{tt2}
Suppose that the copula function  $C(u,v)$ has bounded second-order partial  derivatives on $(0, 1)^2$ and the function $\phi$ admits a bounded derivative $\phi'$. Then, for any sequence of positive constants $(b_n)_{n\geq 1}$ satisfying $0<b_n<1$,\\ $\sqrt{n}b_n^2/\sqrt{\log\log n}=o(1),$ and for some $c>0$, we have almost surely
\begin{equation} \label{biais}
R_n\sup_{\frac{c\log n}{n}\le h\le b_n}\sup_{(u,v)\in(0,1)^2}\vert \mathbb{E}\hat{C}_{n,h}^{(T)}(u,v) - C(u,v)\vert \rightarrow 0,\,\, n\rightarrow\infty.
\end{equation}
\end{theorem}

\begin{remark}{\rm
We can infer from Theorem \ref{tt1} that for any data-driven bandwidth $\hat{h}_n$ such that 
\begin{equation} \label{ee4'}
 \mathbb{P}(\frac{c\log n}{n}\le \hat{h}_n\le b_n)\rightarrow 1,\; n\rightarrow\infty,
 \end{equation}
we have
\begin{equation} \label{ee5}
\sup_{(u,v)\in(0,1)^2}\frac{R_n}{A(c)}\left|\hat{C}_{n,\hat{h}_n}^{(T)}(u,v)-\mathbb{E}\hat{C}_{n,\hat{h}_n}^{(T)}(u,v)\right|\stackrel{\mathbb{P}}{\longrightarrow}1,\ \ \ n\rightarrow \infty,
\end{equation}
where $\stackrel{\mathbb{P}}{\longrightarrow} $ stands for convergence in probability. To make use of \eqref{ee5} for providing confidence bands, we must ensure that the bias of the estimator may be neglected, in the sense that,
\begin{equation}\label{ee6}
\sup_{(u,v)\in(0,1)^2}\frac{R_n}{A(c)}\left|\mathbb{E}\hat{C}_{n,\hat{h}_n}^{(T)}(u,v)-C(u,v)\right|\stackrel{\mathbb{P}}{\longrightarrow}0,\ \ \ n\rightarrow \infty.
\end{equation}
}
\end{remark}

\begin{theorem}\label{tt3}
Suppose that assumptions of Theorem \ref{tt1} and Theorem \ref{tt2} hold and condition \eqref{ee4'} is fulfilled. 
Then, for every $\epsilon \in (0,1)$, as $n\rightarrow\infty$, one has
\begin{multline}\label{lleq1}
\mathbb{P}\bigg(C(u,v)\in \left[\hat{C}_{n,\hat{h}_n}^{(T)}(u,v)-E_{n,\epsilon}(u,v),\hat{C}_{n,\hat{h}_n}^{(T)}(u,v)+E_{n,\epsilon}(u,v)\right],\forall\:0\leq u,v\leq 1 \bigg)\longrightarrow 1,
\end{multline}

\begin{multline}\label{lleq2}
\mathbb{P}\bigg(C(u,v)\in \left[\hat{C}_{n,\hat{h}_n}^{(T)}(u,v)-\Delta_{n,\epsilon}(u,v),\hat{C}_{n,\hat{h}_n}^{(T)}(u,v)+\Delta_{n,\epsilon}(u,v)\right], \forall\:0\leq u,v\leq 1 \bigg)\longrightarrow 0,
\end{multline}
where $E_{n,\epsilon}(u,v)=(1+\epsilon)\frac{A(c)}{R_n},\;\Delta_{n,\epsilon}(u,v)=(1-\epsilon)\frac{A(c)}{R_n}$.
\end{theorem}

 The proofs of our theorems are direct adaptations of those of the results on the Local linear estimator established in \cite{r1}. So we omit the proofs and suggest the interested reader to follow the lines from page 2084 to 2089 in this paper. Instead, we will focus in the next subsection  on simulation studies to compare the performance of our results with the confidence bands obtained from the classical asymptotic normality approach.
\begin{remark} {\rm
Whenever (\ref{lleq1}) and (\ref{lleq2}) hold jointly for every $\epsilon > 0$, we will say that the intervals 
\begin{equation}\label{iinterv}
\left[\alpha_n(u,v),\beta_n(u,v)\right]= \left[\hat{C}_{n,\hat{h}_n}^{(T)}(u,v)-\frac{A(c)}{R_n}\:,\:\hat{C}_{n,\hat{h}_n}^{(T)}(u,v)+\frac{A(c)}{R_n}\right]
\end{equation}
provide asymptotic simultaneous confidence bands for the copula  function $C(u,v),\ 0\leq u,v\leq 1$. These bands are optimal as their asymptotic confidence level tends to 100\%. Therefore we can write, with a probability near to 1,  that for all $(u,v)\in [0,1]^2,$ as $n\rightarrow\infty$,
\begin{equation*}
 C(u,v)\in\left[\alpha_n(u,v),\beta_n(u,v)\right].
\end{equation*}
}
\end{remark}
\subsection{Simulation study}
Here, we make some simulation experiments to show the finite sample performance of our confidence bands. To this end, we choose Frank copula family which has bounded second-order partial derivatives. To compute the estimator $\hat{C}_{n,\hat{h}}^{(T)}$, we  employ the conditional sampling method and generate a random sample of $n$ pairs of data from Frank copula, say $C_{\theta}$, of parameter $\theta\in \mathbb{R}$, defined as 
\begin{equation}
C_{\theta}(u,v)= -\frac{1}{\theta}\log\left[ 1+ \frac{(e^{-\theta u}-1)(e^{-\theta v}-1)}{(e^{-\theta}-1)} \right].
\end{equation}
To control the behavior of the data-driven bandwidth $\hat{h}_n$, we take it close to a sequence of constants $h_n=1/\log(n)$. While the kernel $K(\cdot)$ is taken as the integral of the Epanechnikov kernel density function
$k(t)=0.75(1-t^2)\mathbb{I}(|t|\leq 1)$. We then compute the lower and upper bounds of the confidence bands established in \eqref{iinterv} by substracting and adding respectively the term $A(c)/R_n$, where $A(c)= 1/2$.\\

To measure the performance of our bands, we first compute the true curve $C_{\theta}$, for specific values of $\theta$. Then, we consider $B=1000$ replications of the experiment and determine the frequency with which the bands cover the true curve $C_{\theta}(u,v)$, at all values $0\leq u,v\leq 1$. This approximates the coverage probability of our confidence bands which is reported in Table \ref{tab1}, for different values of $\theta$ and sample size $n=50,100,500$.\\
\begin{table}[htbp]
\begin{center}
$$\begin{array}{ccccc}
\hline
& \text{sample size}\, n &  \theta=-2 &  \theta=1  & \theta=10\\
 \hline
 \multirow{3}*{The proposed method}
& 50 &  0.95  &   0.96 & 0.93\\
& 100 &  0.96  &  0.97  &0.94\\
& 500&  0.98  &  0.99  &0.99\\
\hline
\multirow{3}*{The Normal approximation method}
& 50 &  0.55  &   0.54 & 0.54\\
& 100 &  0.63  &  0.62  & 0.56\\
& 500&  0.60  &  0.66  & 0.62\\
\hline
\end{array}
$$
\end{center}
\caption{Coverage probabilities of the asymptotic confidence bands based on our method and the Normal approximation method.}
\label{tab1}
\end{table}

We can observe that the coverage probability is increasing with $n$ and is satisfactory even for enough small sample sizes, as $n=50$, and is close to 1, for sample sizes reaching 500.\\

We now compare our proposal to the $100(1-\alpha)\%$ confidence bands based on the asymptotic normality of the estimator. Indeed, Omelka et \textit{al.} \cite{r6} (2009) have established, under some regularity assumptions which are fulfilled here by taking $\phi$ equal to the Probit transformation, the weak convergence of the normalized  process $\sqrt{n}[\hat{C}_{n,\hat{h}_n}^{(T)}(\cdot,\cdot) - C(\cdot,\cdot)]$ to a Gaussian limit process with explicit covariance function. This implies that, for any fixed $0\leq u,v\leq 1$, $\sqrt{n}[\hat{C}_{n,\hat{h}_n}^{(T)}(u,v) - C(u,v)]$ converges in distribution to Gaussian random variable with asymptotic variance, 
\begin{eqnarray*}
\sigma^2(u,v)=  C(u,v)[1-C(u,v)-2\{(1-u)C_u(u,v)-(1-v)C_v(u,v)+C_u(u,v)C_v(u,v)\}]&\\
 +\; u(1-u)C^2_u(u,v)+v(1-v)C^2_v(u,v)-2uv C_u(u,v)C_v(u,v),&
\end{eqnarray*}
where $C_u(u,v)$ and $C_v(u,v)$ are the first-order partial derivatives of the copula $C(u,v)$.
Taking $C=C_{\theta}$ representing the Frank copula, we get the explicit expression of $\sigma^2(u,v)$ :
$$ \sigma^2(u,v)=\alpha(u,v)+\beta(u,v),$$
where
\begin{eqnarray*}
\alpha(u,v)& = & C_{\theta}(u,v)\left[ 1-C_{\theta}(u,v)-\frac{2(1-u)e^{-\theta u}(e^{-\theta v}-1)+2(1-v)e^{-\theta v}(e^{-\theta u}-1)}{e^{-\theta(u+ v)}-e^{-\theta u}-e^{-\theta v}+e^{-\theta}} \right]\\
& & + C_{\theta}(u,v)\left[\frac{ 2e^{-\theta(u+ v)}(e^{-\theta u}-1)(e^{-\theta v}-1)}{(e^{-\theta(u+ v)}-e^{-\theta u}-e^{-\theta v}+e^{-\theta})^2}  \right]
\end{eqnarray*}
and
\begin{eqnarray*}
\beta(u,v) &= &\frac{u(1-u)e^{-2\theta u}(e^{-\theta v}-1)^2+v(1-v)e^{-2\theta v}(e^{-\theta u}-1)^2}{(e^{-\theta(u+ v)}-e^{-\theta u}-e^{-\theta v}+e^{-\theta})^2}\\
 &  & + \frac{2uv e^{-\theta(u+ v)}(e^{-\theta u}-1)(e^{-\theta v}-1)}{(e^{-\theta(u+ v)}-e^{-\theta u}-e^{-\theta v}+e^{-\theta})^2}.
\end{eqnarray*}
To compare our method to the normal asymptotic approximation method, we  gives the coverage probabilities of the $99\%$ confidence bands for the true Frank copula $C_{\theta}$ in the last three lines in Table \ref{tab1}. From this table, it is clear that our proposed asymptotic confidence bands are more accurate than those  obtained from the normal asymptotic approximation.\\

\begin{center}
The authors would like to thank the World Bank center for excellence - CEA-MITIC - based at the Universit\'e Gaston Berger for financial support of this work.
\end{center}

\noindent ($\dag$). LERSTAD, Universit\'e Gaston Berger, Saint-louis, SENEGAL.\\
\noindent ($\dag\dag$). LERSTAD, Universit\'e Alioune Diop, Bambey, SENEGAL; cheikhtidiane.seck@uadb.edu.sn.\\
\noindent ($\dag\dag\dag$). LERSTAD, Universit\'e Gaston Berger, Saint-louis, SENEGAL\& LSTA, Universit\'e Pierre et Marie Curie, FRANCE.\\


\addcontentsline{toc}{section}{References}
\begin{thebibliography}{99}
\bibitem{r0} Altman, N. and L\'eger, C. (1995). Bandwidth selection for Kernel distribution function estimation.\textit{ Journal of Statistical Planning and Inference}, 46(2):195-214. DOI: 10.1016/0378-3758(94)00102-2.
\bibitem{r1} B\^a, D., Seck, C.T. and L\^o, G.S. (2015). Asymptotic Confidence Bands for Copulas Based on the Local
Linear Kernel Estimator. \textit{Applied Mathematics}, 6, 2077-2095. http://dx.doi.org/10.4236/am.2015.612183

\bibitem{r2} Deheuvels, P. and Mason, D. M. (2004). General Asymptotic Confidence Bands Based on Kernel-type Function Estimators. \textit{Statistical Inference stochastic process.}, 7:225-277.doi:10.1023/B:SISP.0000049092.55534.af.

\bibitem{r3} Fermanian, J., Radulovic, D. and Wegkamp, M. (2004). Weak convergence of empirical copula processes. \textit{International Statistical Institute (ISI) and Bernoulli Society for Mathematical Statistics and Probability.}, vol. 10, 5:847-860.doi:10.3150/bj/1099579158.

\bibitem{r4} Mason, D., M. and Swanepoel (2010). A general result on the uniform in bandwidth consistency of kernel-type function estimators. \textit{Sociedadde Estadistica e Investigation Operativa 2010.}, doi:10.1007/S11749-010-0188-0

\bibitem{r5} Marron, J. S. and Ruppert, D. (1994). Transformations to reduce boundary bias in kernel
density estimation. Journal of the Royal Statistical Society. Series B (Methodological), 56(4):653-671. DOI: 10.2307/2346189. 

\bibitem{r6} Omelka, M. and Gijbels, I. and Veraverbeke, N. (2009). Improved kernel estimators of copulas : weak convergence and goodness-of-fit testing. \textit{The Annals of Statistics.}, vol. 37, 5B:3023-3058. doi:10.1214/08-AOS666.

\bibitem{r7} van der Vaart, A. W. and Wellner, J. A. Weak Convergence and Empirical Processes, \textit{Springer, New
York}, 1996. doi:10.1007/978-1-4757-2545-2.


\bibitem{r9} Zari, T. (2010). Contribution \`a l'\'etude du processus empirique de copule.  Th\`ese, Mathématiques [math]. Universit\'e Pierre et Marie Curie - Paris VI, 2010. Francais. <tel-00485020>, HAL Id : tel-00485020, version 1.
\end{thebibliography}
\end{document}